\documentclass[useAMS,usenatbib]{mn2e}
\usepackage{txfonts}
\usepackage{graphicx}
\usepackage{amssymb}
\usepackage[below]{placeins}
\usepackage{txfonts}
\usepackage{natbib}
\bibpunct{(}{)}{;}{a}{}{,}
\def \arcmin{$^{\prime}$}

\def \xmm{{\emph{XMM-Newton}}}
\def \rosat{{\emph{ROSAT}}}
\def \chandra{{\emph{Chandra}}}

\newcommand{\ion}[2]{#1\,{\sc{#2}}}

\bibpunct{(}{)}{;}{a}{}{,}

\title[AGN heating and gas uplift in M~87]
{Feedback under the microscope II: heating, gas uplift, and mixing in the nearest cluster core}
\author[N. Werner et al.]{N. Werner$^1$\thanks{Chandra/Einstein fellow, E-mail: norbertw@stanford.edu}, A. Simionescu$^1$\thanks{Einstein fellow}, E. T. Million$^1$, S. W. Allen$^1$, P.~E.~J.~Nulsen$^{2}$,
\newauthor A. von der Linden$^1$, S. M. Hansen$^3$, H.~B{\"o}hringer$^4$, E. Churazov$^{5,6}$, A.~C.~Fabian$^7$,   
\newauthor W.~R. Forman$^2$, C.~Jones$^{2}$, J.~S.~Sanders$^7$, and G.~B.~Taylor$^8$\\
$^1$Kavli Institute for Particle Astrophysics and Cosmology, Stanford University, 382 Via Pueblo Mall, Stanford, CA 94305-4060, USA \\
and SLAC National Accelerator Laboratory, 2575 Sand Hill Road, Menlo Park, CA 94025, USA \\ 
$^{2}$Harvard-Smithsonian Center for Astrophysics, 60 Garden St., Cambridge, MA 02138, USA\\
$^{3}$University of California Observatories \& Department of Astronomy, University of California, Santa Cruz, CA 95064, USA \\
$^4$Max-Planck-Institut f\"ur extraterrestrische Physik, Giessenbachstr, 85748 Garching, Germany\\
$^{5}$Max-Planck-Institut f\"ur Astrophysik, Karl-Schwarzschild-Strasse 1, 85741 Garching, Germany \\
$^{6}$Space Research Institute (IKI), Profsoyznaya 84/32, Moscow 117810, Russia \\
$^{7}$Institute of Astronomy, Madingley Road, Cambridge CB3 0HA \\
$^{8}$Department of Physics and Astronomy, University of New Mexico, Albuquerque, NM 87131, USA\\
}
\begin{document}
\maketitle
\begin{abstract}
Using a combination of deep (574~ks) \chandra\ data, \xmm\ high-resolution spectra, and optical H$\alpha$+[\ion{N}{ii}] images, we study the nature and spatial distribution of the multiphase plasma in M~87. 
Our results provide direct observational evidence of `radio mode' AGN feedback in action, stripping the central galaxy of its lowest entropy gas and therefore preventing star-formation. This low entropy gas was entrained with and uplifted by the buoyantly rising relativistic plasma, forming long ``arms''. A number of arguments suggest that these arms are oriented within $15^{\circ}$--$30^{\circ}$ of our line-of-sight.  The mass of the uplifted gas in the arms is comparable to the gas mass in the approximately spherically symmetric 3.8~kpc core, demonstrating that the AGN has a profound effect on its immediate surroundings. 
The coolest X-ray emitting gas in M~87 has a temperature of $\sim$0.5~keV and is spatially coincident with H$\alpha$+[\ion{N}{ii}] nebulae, forming a multiphase medium where the cooler gas phases are arranged in 
magnetized filaments. We place strong upper limits of 0.06~$M_{\odot}$/yr (at 95 per cent confidence) on the amount of plasma cooling radiatively from 0.5 to 0.25~keV and show that a uniform, volume-averaged heating mechanism could not be preventing the cool gas from further cooling. All of the bright H$\alpha$ filaments in M~87 appear in the downstream region of the $<3$~Myr old shock front, at smaller radii than $\sim$0.6\arcmin. We suggest that shocks induce shearing around the filaments, thereby promoting mixing of the cold gas with the ambient hot ICM via instabilities. By bringing hot thermal particles into contact with the cool, line-emitting gas, mixing can supply the power and ionizing particles needed to explain the observed optical spectra. Furthermore, mixing of the coolest X-ray emitting plasma with the cold optical line emitting filamentary gas promotes efficient conduction between the two phases, allowing non-radiative cooling which could explain the lack of X-ray gas with temperatures under 0.5~keV. 

\end{abstract}


\section{Introduction}

If the energy radiated away at the centres of so-called ``cool-core'' clusters of galaxies, which show sharp X-ray surface brightness peaks and central temperature dips, came only from the thermal energy of the hot, 
diffuse intra-cluster medium (ICM), the ICM would cool and form stars at rates orders of magnitude above what the observations suggest \citep[see][for a review]{peterson2006}. It is currently believed that the energy 
which offsets the cooling is provided by the interaction between the active galactic nuclei (AGN) in the central dominant galaxies and the ICM \citep[see][for a review]{churazov2000,churazov2001,churazov2002,mcnamara2007}. Through a tight feedback loop, it is thought that the AGN can provide enough energy to prevent catastrophic cooling. Recent deep observations of nearby bright cooling core clusters such as Perseus, M~87, Centaurus, and Hydra~A revealed AGN induced weak shocks and sound waves with sufficient energy flux to counterbalance the radiative cooling \citep{sanders2007,forman2005,sanders2008b,nulsen2005,simionescu2009a}. To the first order it is thus understood where the energy comes from and how it gets transported. However, the detailed physics of the AGN/ICM interaction is not yet clear. 

M~87, the central dominant galaxy of the Virgo cluster, at a distance of only 16.1~Mpc \citep{tonry2001} is the nearest, X-ray bright cool core. It is an ideal system for detailed studies of the energy input from the AGN to the hot, 
cooling ICM. 
After the Perseus cluster, M~87 is the second brightest extended extragalactic object in soft X-rays. 
The most striking X-ray features in M~87 are its two ``X-ray arms'', first detected by \citet{feigelson1987} using {\it Einstein Observatory} data and later studied in detail by \citet{boehringer1995} using \rosat\ and VLA 
radio observations. The X-ray arms extend to the East and Southwest from the centre of the galaxy and they spatially correlate with the radio emission. \citet{boehringer1995} found that the X-ray arms are due to 
cooler gas, possibly uplifted from the centre of the galaxy. These results and the subsequent high-quality radio data by \citet{owen2000} led \citet{churazov2001} to argue that the X-ray and radio morphology can be 
explained by bubbles of radio-emitting plasma rising buoyantly through the hot ICM. These rising bubbles entrain and uplift cooler gas from the centre of the galaxy, which then further adiabatically cools. \citet{churazov2001} suggested that the buoyantly rising bubbles dissipate energy into sound waves, internal waves, turbulent motion in the wake, and potential energy of the uplifted gas, all of which could provide heating to the cooling core region 
\citep[see also e.g.][]{bruggen2002,kaiser2003,bruggen2003,deyoung2003,ruszkowski2004a,ruszkowski2004b,heinz2005}. 

Observations with \chandra\ and \xmm\ greatly enhanced our view of M~87. Using \xmm\ observations, \citet{belsole2001}, \citet{matsushita2002}, and \citet{molendi2002} showed that the regions associated with radio arms
have two-temperature (in the ranges of $kT\sim0.8$--1~keV and $kT\sim1.6$--2.5~keV) or multi-temperature structure. However, plasma below 0.8~keV is largely absent \citep{bohringer2002,sakelliou2002}. 
\citet{young2002} analyzed an early, relatively short (40~ks) \chandra\ observation which showed cavities and edges in the hot plasma. They compared the X-ray morphology with the 6~cm radio \citep{hines1989} 
and H$\alpha$+[\ion{N}{ii}] emission \citep{sparks1993} and showed that the optical filaments lie outside of X-ray cavities filled with radio plasma; some are along the edges of the cavities, and several optical 
filaments coincide with knots of cooler X-ray gas. Using a longer ($\sim$150~ks) \chandra\ observation, \citet{sparks2004} studied the relation of the X-ray and optical filaments and concluded that electron 
conduction from the hot X-ray emitting plasma to the cooler phase provides a quantitatively acceptable energy source for the optical filaments. Using the same \chandra\ data, \citet{forman2005} identified shock 
fronts associated with an AGN outburst about 1--2$\times10^7$ yr ago. They argued that shock fronts may be the most significant channel for heating (i.e., entropy increase) of the ICM near to the AGN.

\begin{figure*}
\begin{minipage}{0.47\textwidth}
\includegraphics[height=9.2cm,clip=t,angle=0.,bb=36 102 577 690]{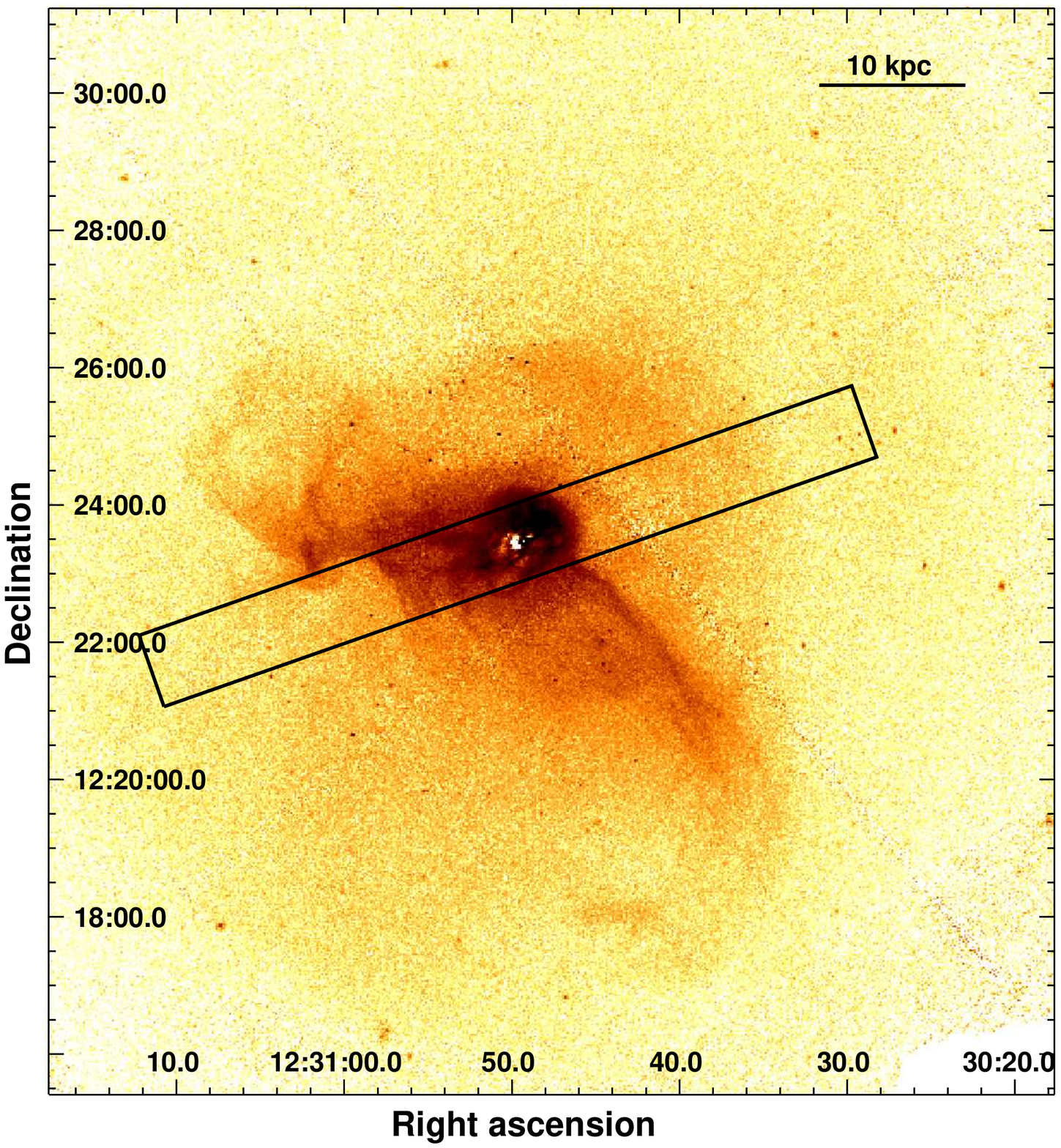}
\end{minipage}
\begin{minipage}{0.47\textwidth}
\includegraphics[height=9.2cm,clip=t,angle=0.,bb=36 102 577 690]{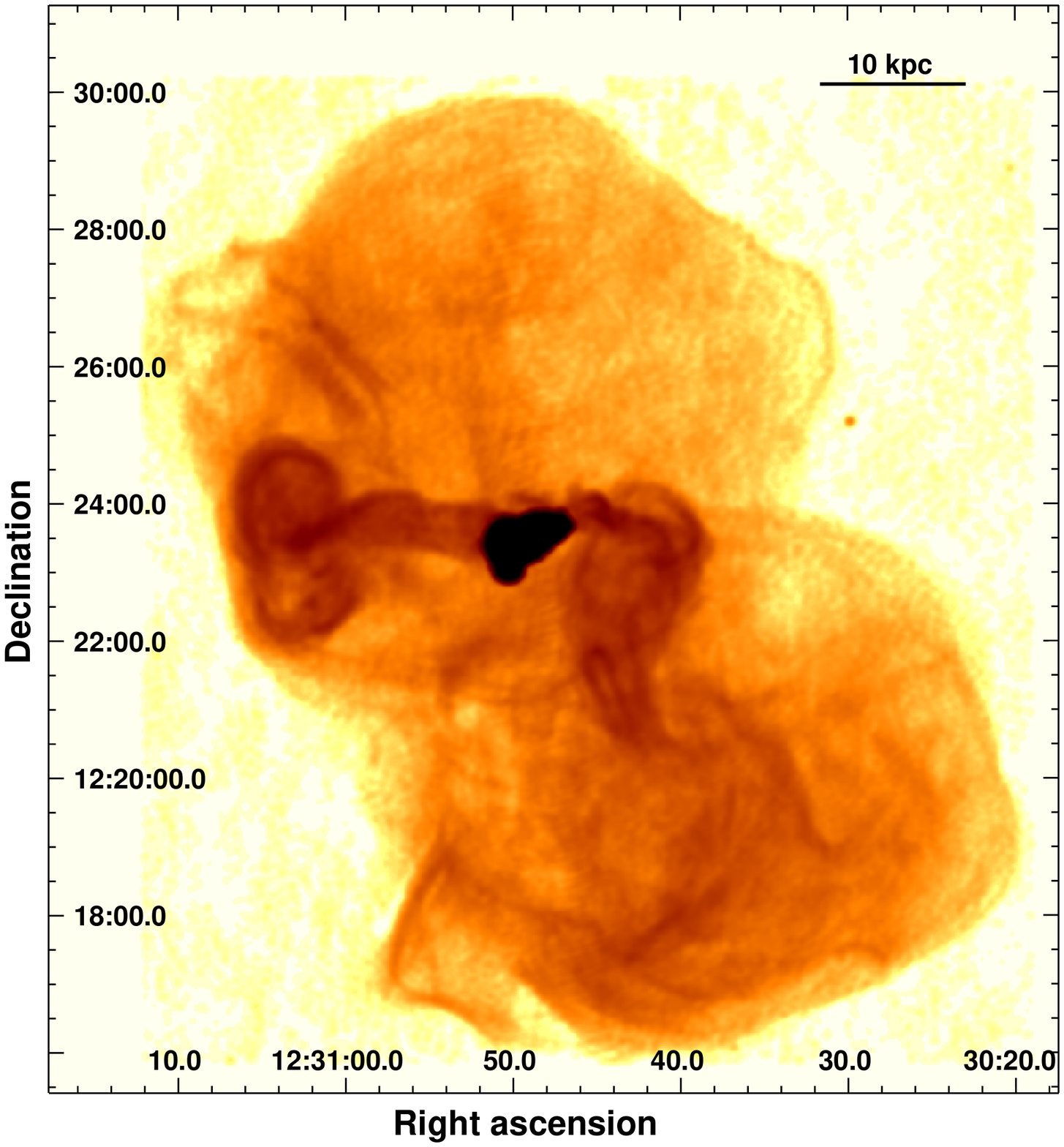}
\end{minipage}
\caption{{\it Left panel: }Relative deviations of the surface brightness from a double beta model fit to the \chandra\ 0.6--2.0~keV image of M~87, obtained by coadding all ACIS-I observations (see Paper~I for details). The cavities in the core and the two X-ray arms are clearly visible. The \xmm\ RGS extraction region is overplotted. {\it Right panel:} The 90~cm radio image by \citet{owen2000}. Two flows (radio arms) emerge from the inner-jet region, one directed Eastward (spatially coincident with the X-ray arm) and the other directed to the Southwest. While the Eastern radio arm ends in edge-brightened torus-like vortex rings, the Southwestern radio arm develops a S-shaped Southward twist. Both radio arms are immersed in a pair of large, partially overlapping radio lobes.  Both panels show the same 14.6\arcmin$\times$15.8\arcmin\ field. } 
\label{M87im}
\end{figure*}

Subsequent \chandra\ observations increased the total exposure time to 500~ks, which allowed \citet{forman2007} to resolve a web of filamentary structure in both X-ray arms. Both arms show narrow filaments, with 
a length-to-width ratio of up to $\sim$50. However, while the Southwestern arm has only a single set of filaments, the Eastern arm shows multiple sets of filaments and bubbles, increasing in scale with the distance from 
the centre. The high spatial resolution of \chandra\ also highlights the fact that while the Eastern X-ray arm is cospatial with the radio arm, the radio emission bends around and avoids the Southwestern X-ray arm. A 
residual image of M~87 showing the inner cavities and the X-ray arms obtained using all \chandra\ ACIS-I data is shown in Fig.~\ref{M87im} \citep[from][]{million2010}, together with the 90~cm radio image by \citet{owen2000}.
The temperature structure and metallicity of the X-ray arms was studied in detail by \citet{simionescu2008a} using deep (120~ks) \xmm\ data. These authors confirmed  that the arms are multiphase with a 
temperature distribution between 0.6--3.2~keV. This temperature distribution is also consistent with that found by spectral fits to the \xmm\ Reflection Grating Spectrometer (RGS) data \citep{werner2006b}, which 
show the presence of weak \ion{Fe}{xvii} lines. \citet{simionescu2008a} estimate the total mass of gas below 1.5~keV, most of which was uplifted by the AGN, as $5\times10^{8}~M_{\odot}$. 

This is the second in a series of papers \citep[following the work by][hereafter Paper I]{million2010} presenting detailed spatially resolved spectroscopy of M~87, exploiting the superb spatial resolution and excellent photon statistics of the deep (574~ks) \chandra\ observation, to study AGN feedback and ICM physics.  In this paper, we study for the first time the detailed nature and the spatial distribution of the multiphase 
plasma associated with the X-ray arms. \chandra\ allows us to study detailed structure in the cooler phases at a much higher spatial resolution than is possible with \xmm. 
Our data allow us to sample well the small substructure seen in the \chandra\ images, map the spatial distribution of the individual temperature components and compare it with new H$\alpha$ images obtained at the {\it Lick Observatory}, and archival data from the {\it Hubble Space Telescope (HST)}. In Section 2, we describe the data reduction and analysis; in Section 3, we present the results of the observations; in Section 4, we determine the physical properties of the cooler plasma phases and discuss the uplift of the low entropy gas from the bottom of the gravitational potential well, and effects of shocks and mixing. Finally, in Section 5, we summarize our conclusions.

A distance to M~87 of 16.1~Mpc \citep{tonry2001} implies a linear scale of 4.7~kpc~arcmin$^{-1}$. All errors are quoted at the 68 per cent confidence level for one interesting parameter.

\begin{table}
\begin{center}
\caption{Summary of the \chandra\ observations.
Colums list the 
observation ID, detector, observation mode, exposure after cleaning, 
and observation date.}\label{table:obs}
\vskip 0 truein
\begin{tabular}{ccccccc}
\hline\hline
Obs. ID & Detector & Mode & Exposure (ks) & Obs. date\\
\hline 
2707 &  ACIS-S & FAINT  & 82.9 & Jul. 6  2002\\
3717 &  ACIS-S & FAINT  & 11.1 & Jul. 5  2002\\
5826 &  ACIS-I & VFAINT & 126.8 & Mar. 3 2005\\
5827 &  ACIS-I & VFAINT & 156.2 & May 5  2005\\
5828 &  ACIS-I & VFAINT & 33.0 & Nov. 17 2005\\
6186 &  ACIS-I & VFAINT & 51.5 & Jan. 31 2005\\
7210 &  ACIS-I & VFAINT & 30.7 & Nov. 16 2005\\
7211 &  ACIS-I & VFAINT & 16.6 & Nov. 16 2005\\
7212 &  ACIS-I & VFAINT & 65.2 & Nov. 14 2005\\
\hline
\label{obs}
\end{tabular}
\end{center}
\end{table}

\section{Data reduction and analysis}

\subsection{{\it Chandra} data}

The \chandra\ observations of M~87 were taken between July 2002 and November 2005 using the Advanced CCD Imaging Spectrometer (ACIS). The observations are listed in Table~\ref{obs}. The total net exposure time after cleaning is 574~ks. We follow the data reduction procedure described in Paper~I \citep[see also][]{million2008,million2009}.

The individual regions for the spectral analysis were determined using the Contour Binning algorithm \citep{sanders2006b}, which groups neighboring pixels of similar surface brightness until a desired signal-to-noise threshold is met. In order to have small enough regions to resolve substructure and still have enough counts to fit a simple multi-temperature model, we adopted a signal-to-noise ratio of 50. The total number 
of regions in this analysis is $\sim$6000, with a total number of $\sim$15,000,000 net counts. 

Background spectra for the appropriate detector regions were extracted from the blank-sky fields available from the Chandra X-ray Center. These were normalized by the ratio of the observed and blank-sky count 
rates in the $9.5-12$ keV band (the statistical uncertainty in the observed $9.5-12$ keV flux is less than 5 per cent in all cases). The background level in these observations is low and represents only a marginal 
source of systematic uncertainty in the determined quantities.

Spectral modeling has been performed with the SPEX package \citep[][SPEX uses an updated version of the MEKAL plasma model with respect to XSPEC]{kaastra1996}.  We analyze data in the 0.6--7.0 keV band. To investigate the multi-temperature structure and map the spatial distribution of plasmas at different temperatures, we fit to each bin a model consisting of collisionally ionized equilibrium plasmas at four fixed temperatures, with variable normalizations and common metallicity. This method was introduced by \citet[][]{sanders2004} in the analysis of deep \chandra\ observations of the Perseus cluster. For M~87, the temperatures are fixed at 0.5, 1.0, 2.0, and 3.0~keV. The temperatures of the coolest and hottest components were selected based on previous multi-temperature analysis of \xmm\ RGS and EPIC data \citep{werner2006b,simionescu2008a}. We also searched for a 0.25~keV component in our spectral fits, but did not detect any emission at this temperature. The temperatures of the individual spectral components are sufficiently far apart that the spectral analysis can constrain their emission measures if they are simultaneously present in the same extraction region. The overall metallicity in each region is free to vary. The O/Fe ratio is fixed at 0.59 Solar, Ne/Fe at 1.20 Solar, and the Mg/Fe at 0.60 Solar, as determined from \xmm\ RGS spectra \citep{werner2006b}. The Galactic absorption toward M~87 is modelled as neutral gas with Solar abundances with a column density fixed at $N_{\mathrm{H}}=1.94\times10^{20}$~cm$^{-2}$, the value determined by the Leiden/Argentine/Bonn (LAB) Survey of Galactic \ion{H}{i} \citep{kalberla2005}. Throughout the paper, abundances are given with respect to the ``proto-solar values'' by \citet{lodders2003}, which are for O, Ne, and Fe approximately 30 per cent lower than the values given by \citet{anders1989}.

\begin{figure*}
\begin{minipage}{0.47\textwidth}
\includegraphics[height=8.3cm,clip=t,angle=0.,bb= 36 123 577 670]{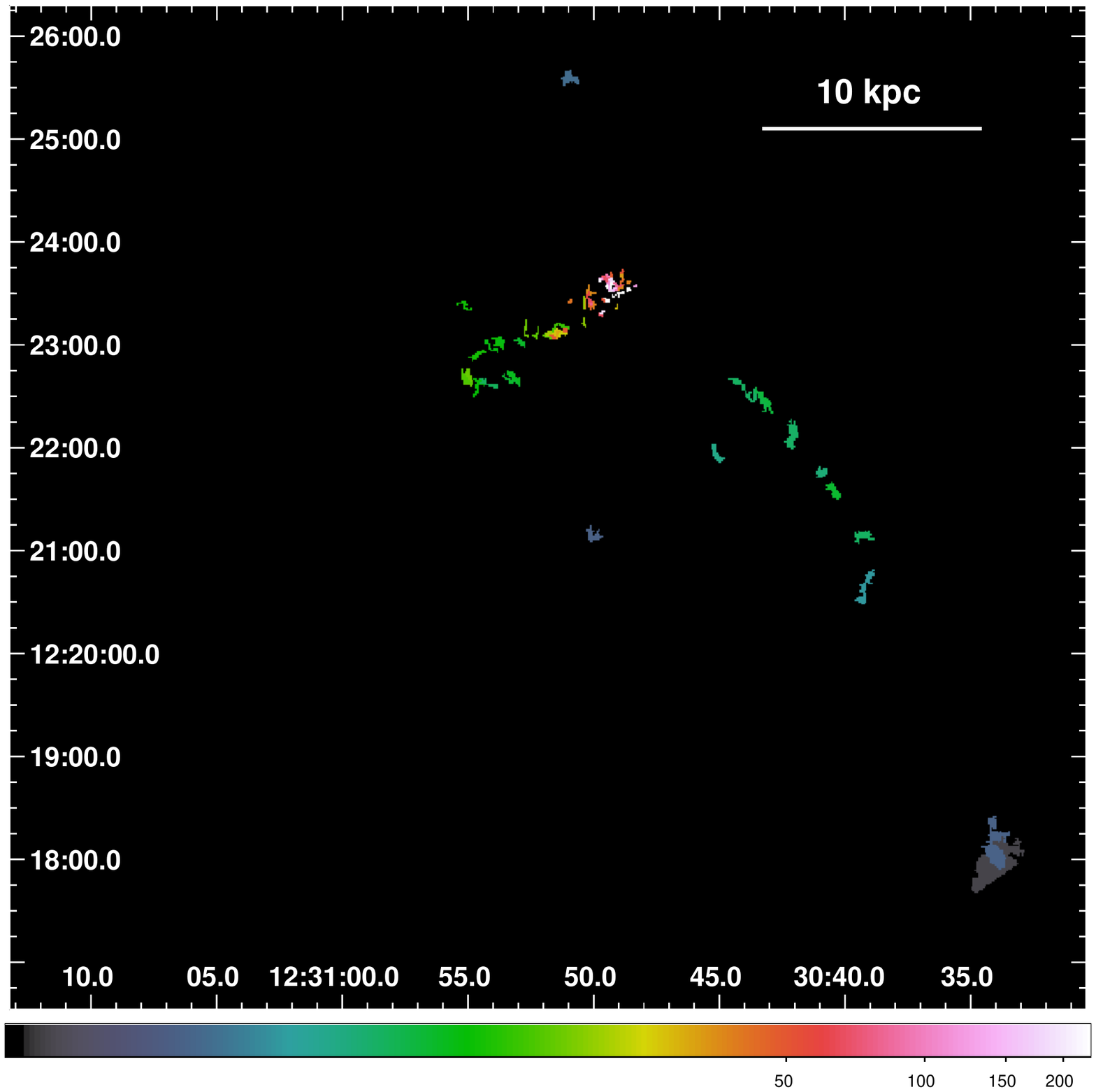}
\end{minipage}
\begin{minipage}{0.47\textwidth}
\includegraphics[height=8.3cm,clip=t,angle=0.,bb= 36 123 577 670]{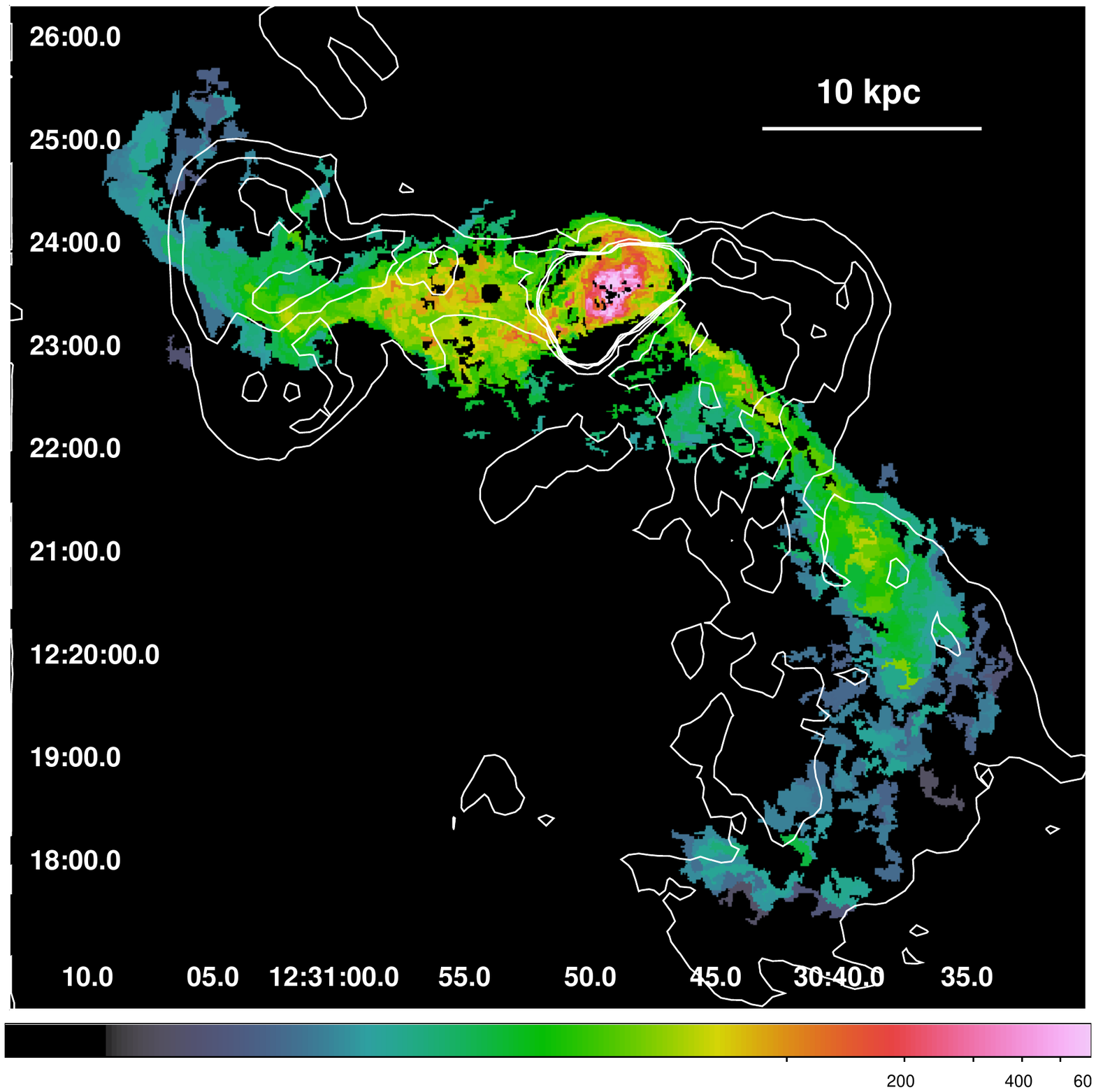}
\end{minipage}\\
\vspace{2mm}

\begin{minipage}{0.47\textwidth}
\includegraphics[height=8.3cm,clip=t,angle=0.,bb= 36 123 577 670]{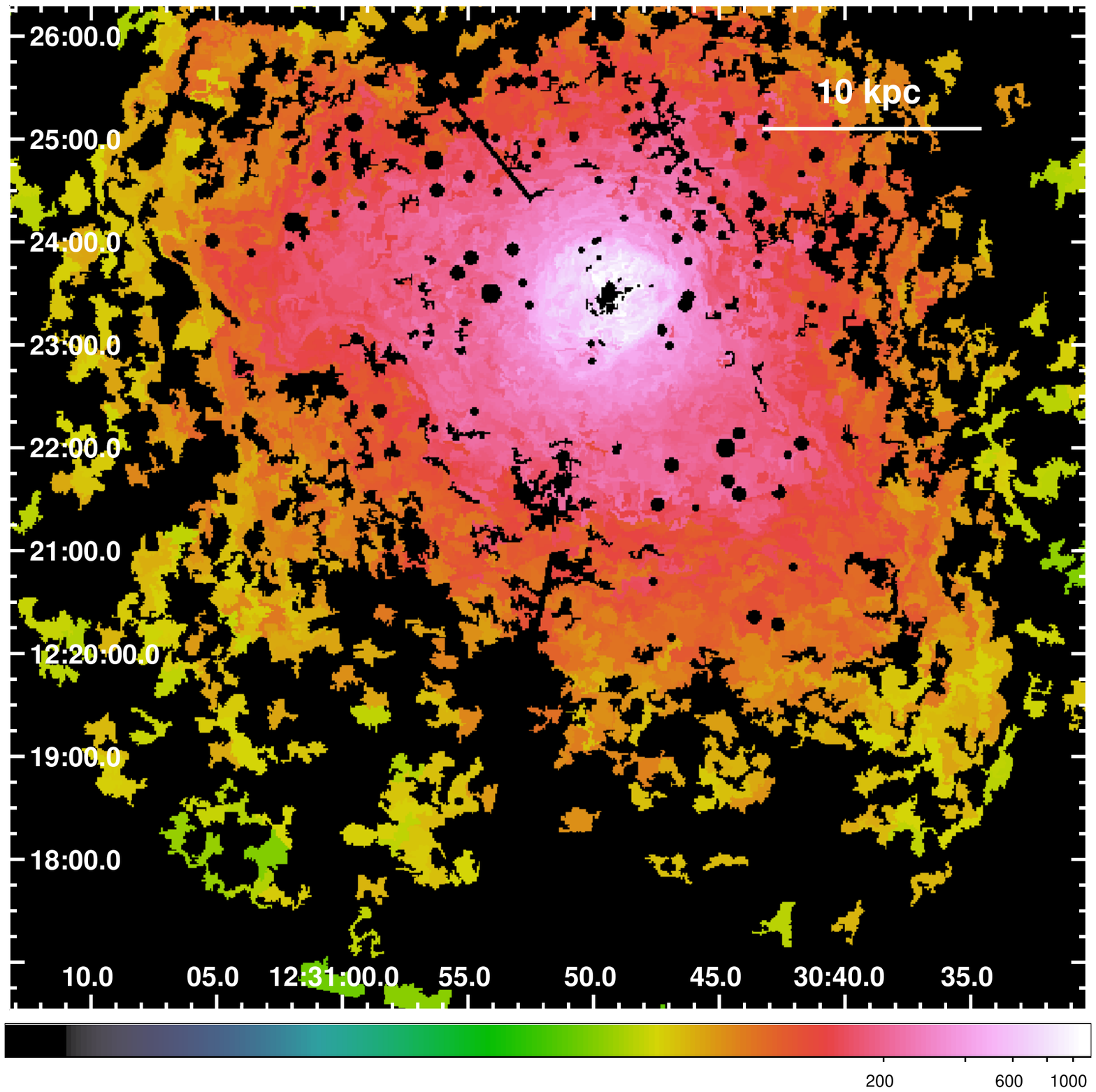}
\end{minipage}
\begin{minipage}{0.47\textwidth}
\includegraphics[height=8.3cm,clip=t,angle=0.,bb= 36 123 577 670]{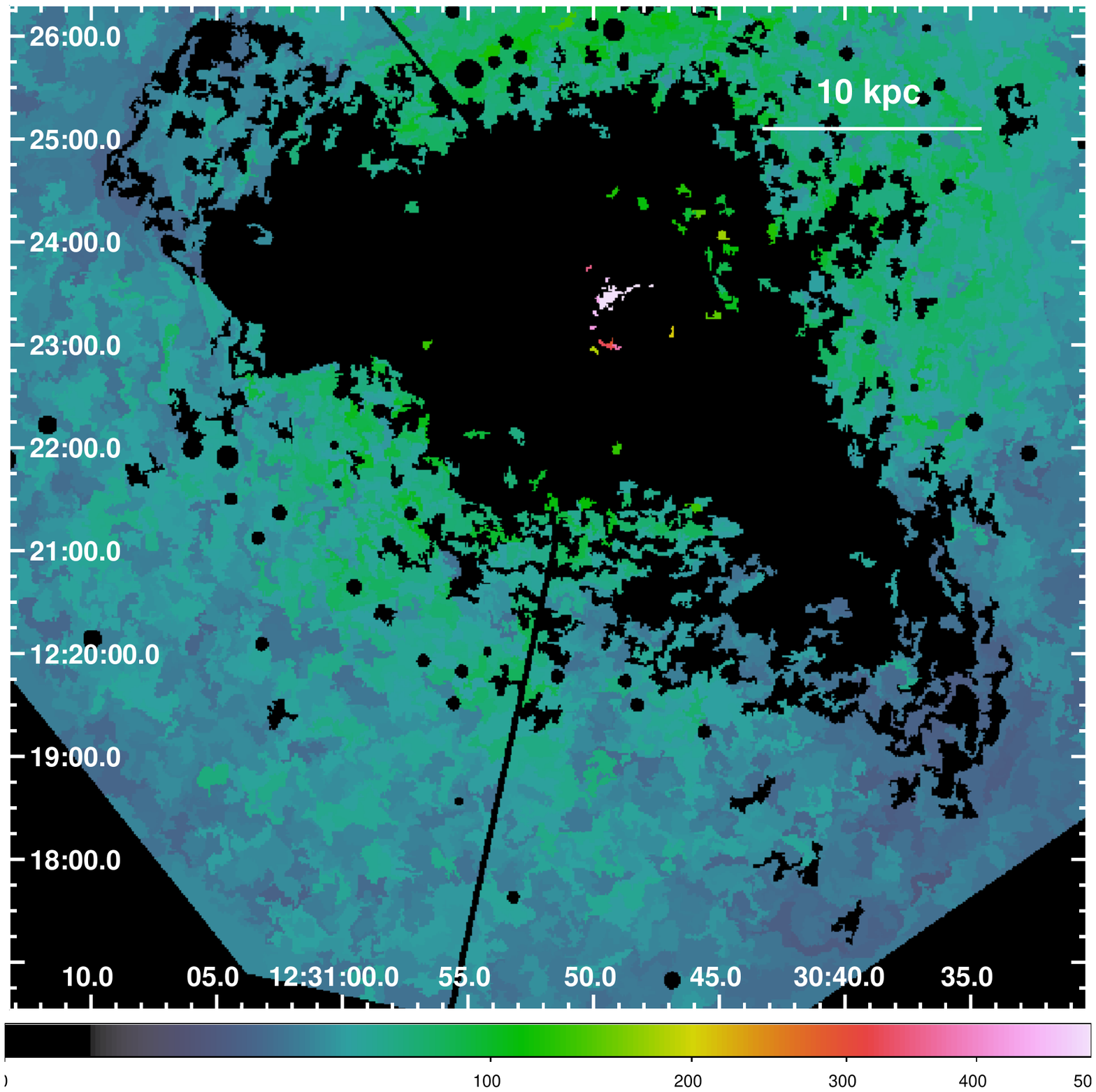}
\end{minipage}
\caption{Spatial distribution of the emission measure, $Y=\int n_{\mathrm{H}}n_{\mathrm{e}}\mathrm{d}V$, (in $10^{58}$~cm$^{-3}$~arcsec$^{-2}$) of the 0.5~keV (upper left), 1.0~keV (upper right), 2.0~keV 
(bottom left), and 3.0~keV (bottom right) plasma detected at 99.7 per cent confidence. On the upper right panel, we over-plotted the contours of the 90~cm radio image \citep{owen2000}. All the panels show the same 
10.45\arcmin$\times$9.75\arcmin\ field. For a zoomed in version of the spatial distribution of 0.5~keV plasma in the core see the right panel of Fig.~\ref{Halpha}.}
\label{maps}
\end{figure*}

\subsection{\xmm\ RGS data}

The multi-temperature structure of the hot plasma in and around M~87 exhibits a complex set of spectral lines that cannot be modelled by a single-temperature spectral model \citep{sakelliou2002,werner2006b}. In 
order to confirm that our simple four-temperature model describes the temperature structure of M~87 well, we fit the same model to the \xmm\ RGS spectra. 

The \xmm\ RGS data were obtained in January 2005, with a net exposure time of 84~ks. The data were processed as described in \citet{werner2006b}. Spectra were extracted from a 1.1\arcmin\ wide extraction 
region, centred at the core of the galaxy. Because the RGS operates without a slit, it collects all photons from within the 1.1\arcmin$\times\sim$12\arcmin\ field of view. Line photons originating at angle $\Delta\theta
$ (in arcminutes) along the dispersion direction are shifted in wavelength by $\Delta\lambda=0.138\Delta\theta$~\AA. Therefore, every line is broadened by the spatial extent of the source. To account for this spatial 
broadening in our spectral model, we produce a predicted line spread function (LSF) by convolving the RGS response with the surface brightness profile of the galaxy derived from the EPIC/MOS1 image along the 
dispersion direction. Because the radial profile of a particular spectral line can be different from the overall radial surface brightness profile, the line profile is multiplied by a scale factor $s$, which is the ratio of the 
observed LSF to the expected LSF. This scale factor is a free parameter in the spectral fit. 

We fit the spectra in the 8--38~\AA\ band with the same four-temperature model fitted to the \chandra\ data, with an additional 0.25~keV component included to determine an upper limit on the amount of such cool 
plasma. We have also separately fitted the RGS data with a model consisting of four isobaric cooling flows with elemental abundances tied between the models \citep[similar to one of the models used for the Centaurus 
cluster in][]{sanders2008}. In this case, we model separately gas cooling from 3~keV to 2~keV, from 2~keV to 1~keV, from 1~keV to 0.5~keV, and from 0.5~keV to 0.25~keV. 
The central AGN cannot be excluded from the RGS data of the core of the galaxy and also needs to be accounted for. 
We model the AGN spectrum with a power-law of a photon index $\gamma=1.95$ and a 2--10 keV luminosity of $L_{\mathrm{X}}=1.15\times10^{42}$~erg~s$^{-1}$ \citep{werner2006b}.

\subsection{H$\alpha$ data}

Narrow-band H$\alpha$ imaging (central wavelength 6606~\AA, FWHM 75~\AA) of M~87 was carried out using the {\it Shane} 3~m telescope at {\it Lick
Observatory} on March 26th, 2009. The total integration time was
24~ks, split into 20 exposures using a dither pattern with $\sim
5$\arcsec\ offsets. We also acquired 1200 seconds of $R$ band imaging,
also split into 20 exposures. The data were processed using the
pipeline of \citet{esm05}. A constant background was subtracted,
estimated from the region of the image with the lowest counts. The
images were registered with {\tt scamp} \citep{ber06}, and resampled
and combined with {\tt swarp} \citep{bmr02}. The seeing of the coadded
images is about 2\arcsec.

To study the morphology of the brighter H$\alpha$ filaments in the cluster
core, we have also analyzed two H$\alpha$ images available in
the {\it HST} archive.  Both images were taken with WFPC2, placing the core
onto the Planetary Camera. The Southeastern region was imaged with the
Wide Field Camera for only 2700 seconds (proposal ID: 5122, PI Ford), and is
not as deep as our ground-based image. The Northwestern region was
imaged significantly longer (13900~seconds, proposal ID: 6296, PI Ford).

\section{Results}

\subsection{Spatial distribution of the X-ray gas}

\begin{figure}
\includegraphics[width=1.15\columnwidth,angle=0.]{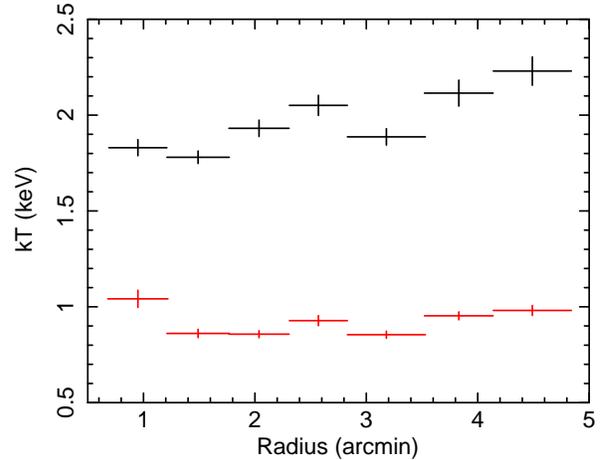}
\caption{The temperature profile of the Southwestern X-arm fitted with a two-temperature model. While the temperature of the hotter component shows a slight radial increase (black data points) the radial distribution of the cooler temperature component (red data points) looks remarkably flat at around 1~keV. If the uplift were adiabatic, the temperature of the cooler gas within the arm would be radially decreasing. } 
\label{kTprof}
\end{figure}

\begin{figure*}
\begin{minipage}{0.47\textwidth}
\includegraphics[width=1.02\columnwidth,clip=t,angle=0.,bb= 36 207 577 585]{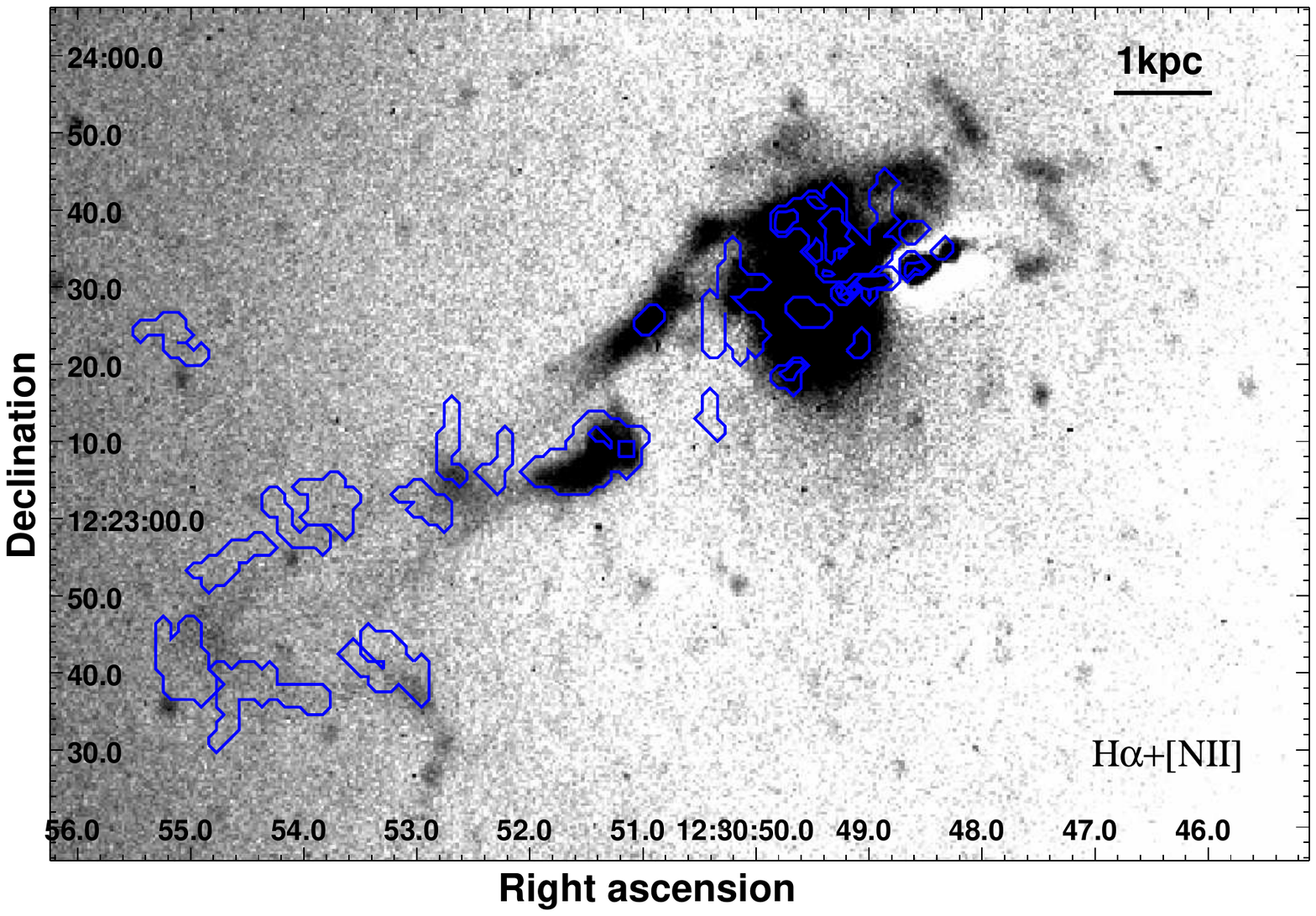}
\end{minipage}
\begin{minipage}{0.47\textwidth}
\includegraphics[width=1.02\columnwidth,clip=t,angle=0.,bb= 36 207 577 585]{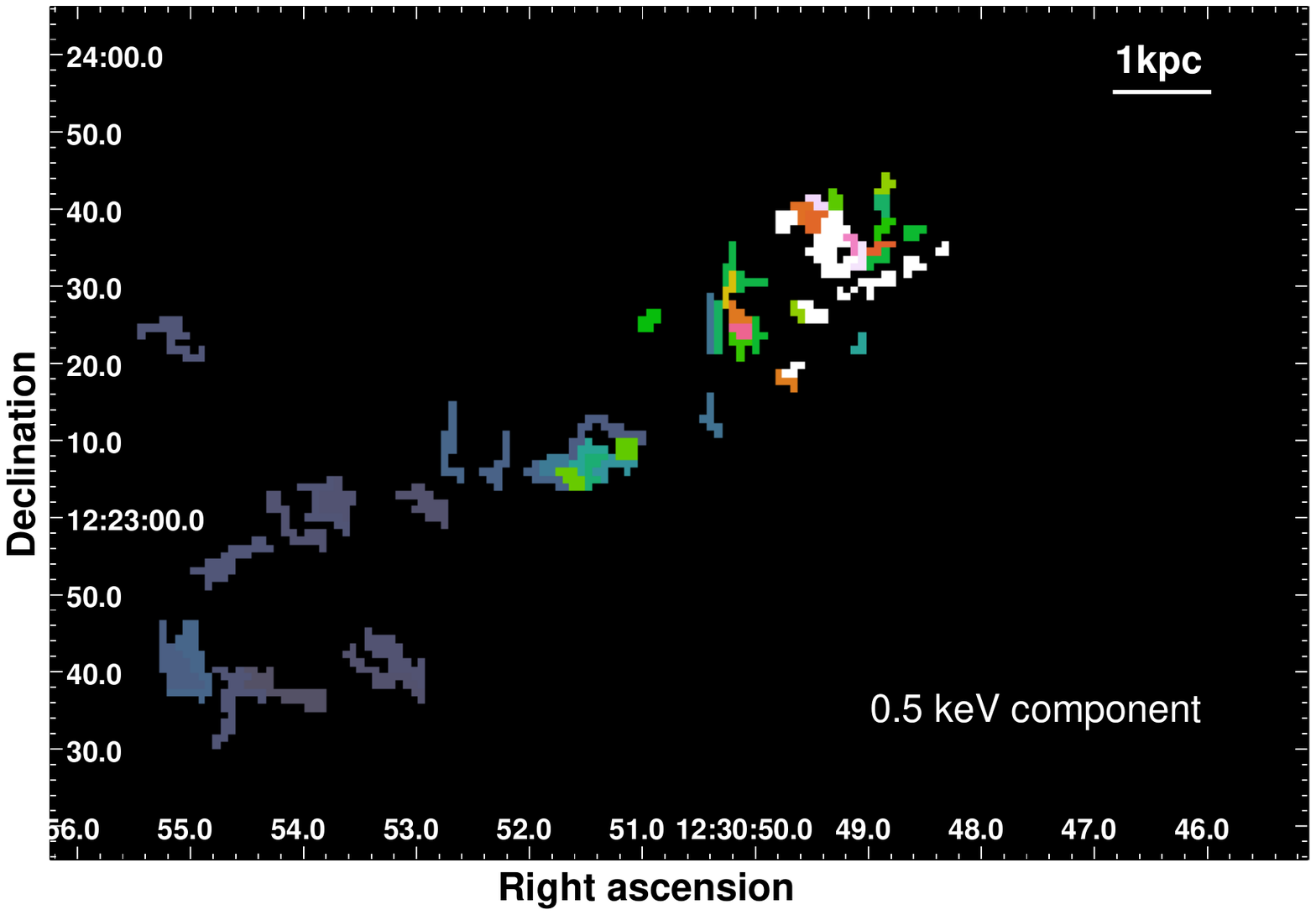}
\end{minipage}\\
\caption{The H$\alpha$ emission (left panel) and plasma at $\sim$0.5~keV (right panel; blue contours in the left panel) spatially correlate in the core and form a horseshoe like feature to the Southeast of the core. The right panel is a zoomed in version of the top left panel of Fig.~\ref{maps}. Both panels show the same 2.7\arcmin$\times$1.85\arcmin\ field.} 
\label{Halpha}
\end{figure*}

\begin{figure*}
\begin{minipage}{0.47\textwidth}
\includegraphics[width=0.97\columnwidth,clip=t,angle=0.,bb= 36 125 577 668]{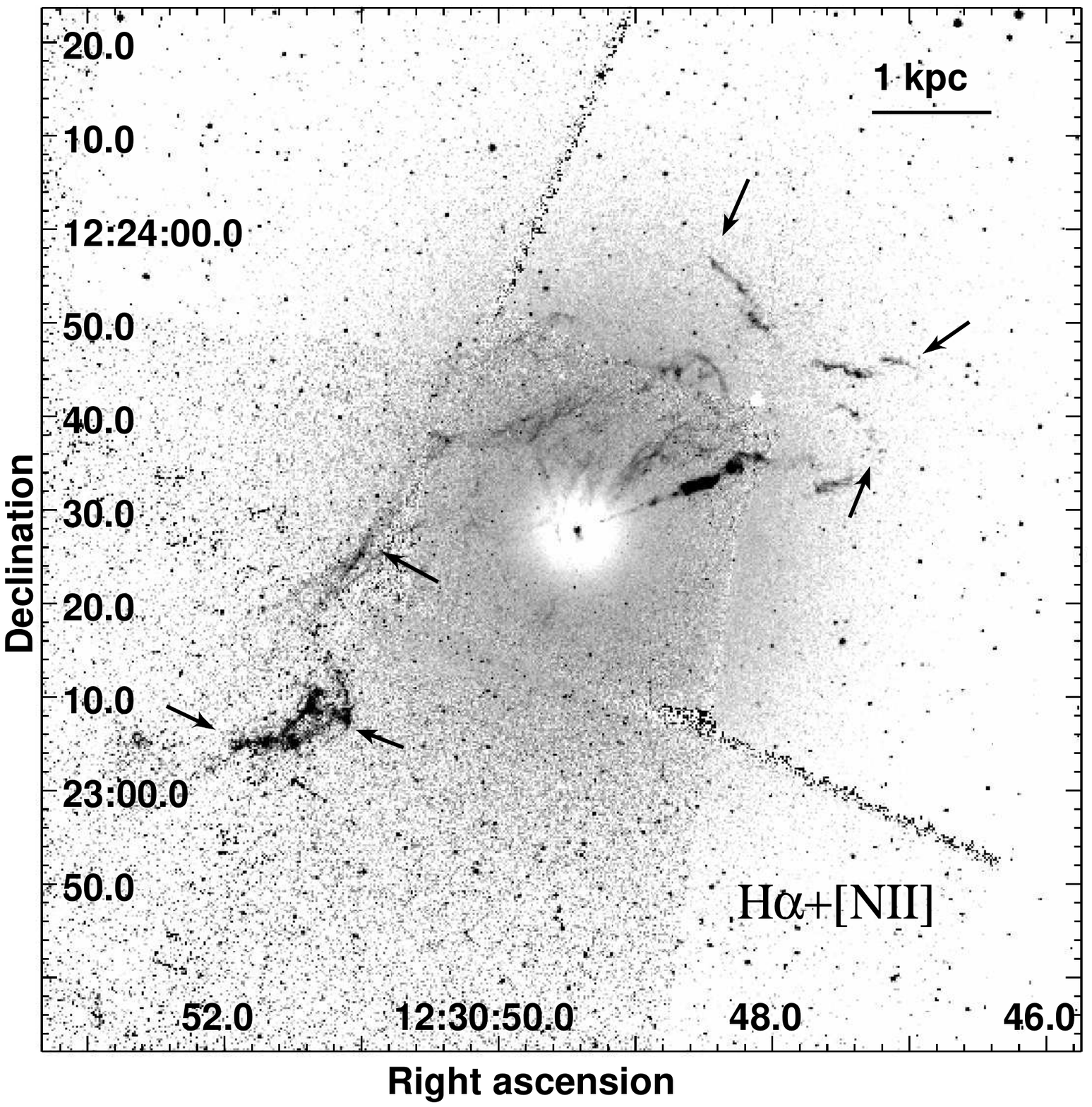}
\end{minipage}
\begin{minipage}{0.47\textwidth}
\includegraphics[width=1.05\columnwidth,clip=t,angle=0.,bb= 36 151 577 641]{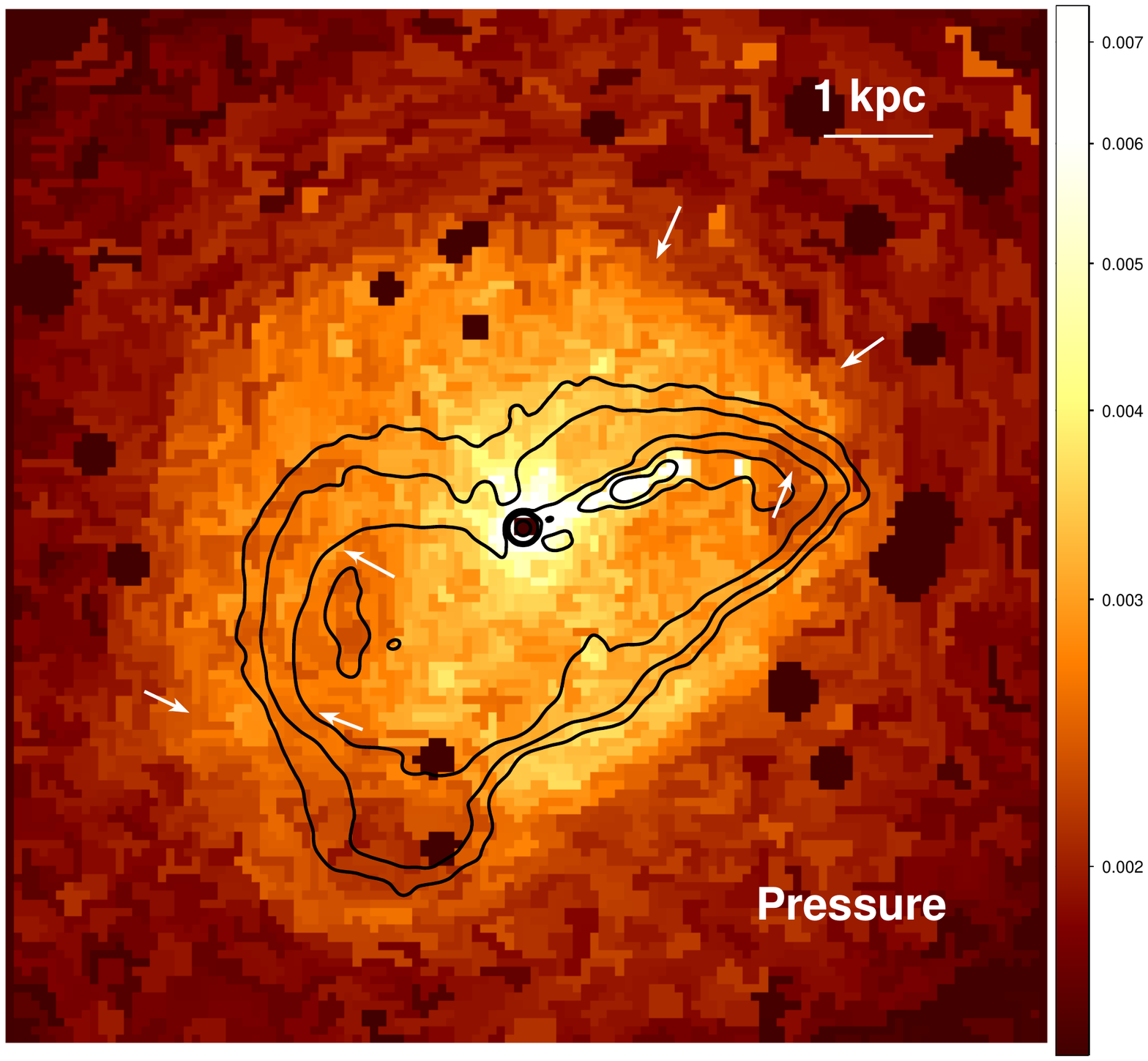}
\end{minipage}\\
\caption{{\it Left panel:} H$\alpha$ image of the innermost 2\arcmin$\times$2\arcmin\ region of M~87, divided by the best fit De Vaucouleurs profile, obtained by {\it HST}. {\it Right panel:} \chandra\ pressure map  
(in units of keV~cm$^{-3}\times\left(\frac{1}{2\mathrm{Mpc}}\right)^{-1/2}$, see Paper~I for details on producing the pressure map) of the same inner 2\arcmin$\times$2\arcmin\ region overplotted with the 6~cm radio emission contours from \citet{hines1989}. The pressure map clearly shows a discontinuity, a likely shock front, at a radius of 0.6\arcmin. The H$\alpha$ image shows that all the bright H$\alpha$ filaments appear in the downstream region of the shock. The arrows identify the bright H$\alpha$ features furthest from the centre, at the edges of apparently underpressured X-ray cavities, and a bright H$\alpha$ ``knee'' on the Southeast.}
\label{P_Halpha}
\end{figure*}

Fig.~\ref{maps} shows the spatial distribution of the emission measures for the four individual temperature components (0.5, 1.0, 2.0, and 3.0~keV) obtained from the \chandra\ data. The top left panel shows 
an interesting result: the map reveals the presence of gas with a temperature of $kT\sim$0.5~keV to the North of the jet in the core of the galaxy, in the Southwestern arm, and in a horseshoe like filament to the Southeast of the core outside of the mushroom shaped Eastern radio arm (radio contours are overplotted in the top right panel of Fig.~\ref{maps}). No 0.5~keV gas with significance higher than 3$\sigma$ is present within the Eastern radio arm. While all the detected 0.5~keV plasma in the core and in the Southeastern horseshoe is within the RGS extraction region (see the left panel of Fig.~\ref{M87im}), the Southwestern arm is outside of the area covered by the present RGS data. 

The spatial distribution of the $\sim$1~keV plasma seen in the top right panel of Fig.~\ref{maps} correlates with the radio emission and is very similar to the morphology of the X-ray arms seen clearly in Fig.~\ref{M87im}. 
In the core, the 1~keV plasma forms a prominent ridge to the North of the AGN, which is cospatial with 0.5~keV plasma and H$\alpha$+[\ion{N}{ii}] emission (see below). 
Based on two-temperature spectral fits to \xmm\ data it has been shown that the X-ray arms are relatively isothermal at $\sim$1~keV \citep{molendi2002,simionescu2008a} and our spectral fits to the Southwestern X-ray arm confirm that its temperature is constant as a function of radius (see Fig.~\ref{kTprof}). 
Most of the Eastern radio arm contains $\sim$1~keV plasma, a large amount of which is also present to the South of the ``stem of the radio mushroom'' where the 0.5~keV horseshoe is observed. 
The brightest part of the Southwestern radio arm is not spatially coincident with the 0.5~keV and 1~keV plasma, but it bends around the cooler X-ray gas. On the 
Southwest, just outside of the core region, the 1~keV plasma forms a narrow and remarkably straight and smooth filament, which broadens with increasing distance from the core. At around 3\arcmin, which is the approximate distance of the circular shock front described in Paper~I  and by \citet{forman2007}, the distribution of 1~keV gas broadens, and at 4.5\arcmin\ it starts to bend to the Southeast.  

The spatial distribution of the plasma at $\sim2$~keV  is, in comparison, remarkably symmetric and only slightly elongated in the direction of the X-ray arms. A knot of $\sim$3~keV plasma is present at the inner 
edge of the X-ray cavity South of the core, at the bottom of the bright radio cocoon, which \citet{forman2005} identified as a ``bud''. This 3~keV plasma is most probably associated with a shock front at the radius of 
0.6\arcmin\ \citep[][Paper~I, see also the right panel of Fig.~\ref{P_Halpha}]{forman2007}.

The fraction of the emission measure of cooler gas (1~keV and 0.5~keV) in the X-ray arms with respect to the total emission measure is 10--35 per cent, consistent with the results obtained by fitting a differential emission measure distribution to \xmm\ data by \citet{simionescu2008a}. The spatial distribution of the coolest gas phases is also in good agreement with that found in the \xmm\ data analysis \citep{simionescu2008a}, but is revealed by \chandra\ with a strikingly better spatial resolution. 

The elemental abundances obtained from the fits are likely to be biased due to the limitations of our model, which has its temperatures fixed at specific values \citep[detailed study of 
the metallicity will be presented in Million et al. in prep.; see also Paper~I and][]{simionescu2008a}.

\subsection{Morphology of the H$\alpha$ nebulae: new clues about their powering}

The left panel of Fig. \ref{Halpha} shows an excerpt of the H$\alpha$+[\ion{N}{ii}] image obtained at the {\it Lick Observatory} after subtraction of the scaled $R$-band continuum image. Owing to the large depth, our 
data reveal a previously undiscovered horseshoe-like filament of H$\alpha$+[\ion{N}{ii}] emission to the Southeast of the core. The feature is detected at a mean level of  5$\sigma$ above the sky. Fig.~\ref{Halpha} shows that the spatial distribution of the H$\alpha$+[\ion{N}{ii}] emission is remarkably similar to that  of the 0.5~keV plasma, both in the core and in the Southeastern horseshoe like region. 
Except for a small knot of H$\alpha$+[\ion{N}{ii}] emission in the ``cap of the mushroom'' \citep[see][]{gavazzi2000}, the Eastern radio arm is devoid of observable H$\alpha$ filaments. As mentioned above, the 
Eastern radio arm is also devoid of any detectable 0.5~keV emission. 
Our {\it Lick Observatory} data (Fig.~\ref{Halpha})  and the {\it HST} data (Fig.~\ref{P_Halpha}) unfortunately do not cover the Southwestern X-ray arm.

Fig.~\ref{P_Halpha} shows an H$\alpha$ image for the innermost 2\arcmin$\times$2\arcmin\ region of M~87 from {\it HST} next to a pressure map obtained in Paper~I by using \chandra\ data. In the pressure map we see a discontinuity, likely shock front, at 
0.6\arcmin\ and several underpressured regions filled with radio emitting plasma, as shown by the contours of 6~cm radio emission from \citet{hines1989}. Assuming that this inner shock propagates at an 
approximately constant velocity of 1000~km~s$^{-1}$, it originated in an AGN outburst $\sim$2.8 Myr ago. Interestingly, all the bright H$\alpha$ emission is within the inner higher pressure region surrounded by the 
shock front at 0.6\arcmin. Some filaments appear to be parallel to the shock front, others seem to lie at the edges of cavities, while the filament to the Southeast of the core is perpendicular to the shock front, bright in 
the downstream region and faint in the upstream. Upstream of the shock front, the Southeastern H$\alpha$ filament continues at a much lower surface brightness and forms the prominent horseshoe.  On the H$\alpha$ image, we identify with arrows the bright features furthest from the centre, at the edges of X-ray cavities, and a bright H$\alpha$ ``knee'' on the Southeast.  The same arrows are plotted in the pressure map.

\subsection{High-resolution \xmm\ RGS spectra}

\begin{figure}
\includegraphics[width=1.1\columnwidth,clip=t,angle=0.]{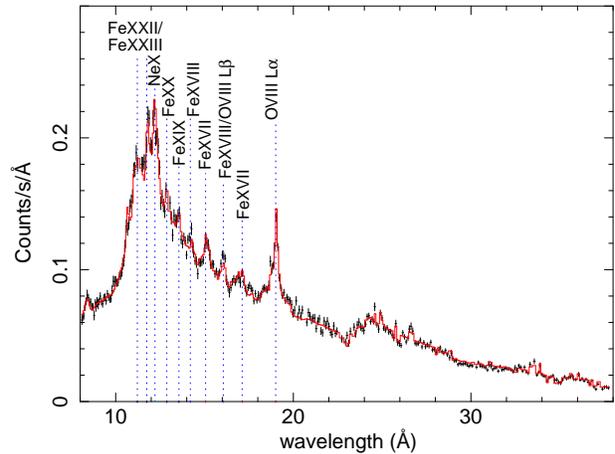}
\caption{The first order RGS spectrum extracted from a 1.1\arcmin\ wide region centred on the core of M~87. The continuous line represents the best fit model to the spectrum. } 
\label{fig:rgs}
\end{figure}

To confirm the presence of the coolest gas phases indicated by fitting the \chandra\ data and to verify that the simple four-temperature model fits the data sufficiently well, we also analyze the \xmm\ RGS high-resolution line spectra. The results of the five-temperature fit to the RGS data are listed in Table~\ref{RGStable} and the best fit model to the spectra is shown in Fig.~\ref{fig:rgs}. 

The emission measures for the 2~keV and 1~keV phases obtained by the RGS are consistent with those from the \chandra\ maps for the area covered by the gratings. The emission measure of the 0.5~keV phase obtained using the RGS is higher by a factor of 3 than the value from the \chandra\ map. The sensitivity of \chandra\ in the 0.6--7~keV band to 0.5~keV emission in these moderate S/N spectra is limited and if we only consider regions where the 0.5 keV component was detected at larger than the 99.7\% confidence level then we miss the cool emission in some regions. By including all regions detected at the 68\% confidence the 0.5~keV features in the map become wider and the agreement between the emission measures from \chandra\ and RGS improves. 

The lack of \ion{O}{vii} lines provides strong constraints on the amount of 0.25~keV plasma. The 95 per cent upper limit for its emission measure is lower by a factor of 11 than the emission measure of the 0.5~keV plasma.

To get constraints on the amount of radiative cooling, we also fit the RGS spectra with a set of cooling flow models (see Table~\ref{RGStable}). If the plasma would cool isobarically in the absence of heating, 
the best fit mass deposition rates would be the same in the different temperature ranges. However, the mass deposition rates are decreasing as a function of temperature and for the gas cooling from 0.5~keV to 
0.25~keV we determine a tight 95 per cent upper limit of $\dot{M}=0.06~M_{\odot}$/yr. 

\begin{table}
\begin{center}
\caption{Best fit parameters for a five-temperature fit and a four-cooling-flow model fit to the high-resolution RGS spectra extracted from a 1.1\arcmin\ wide region centred on the core of M~87. Emission measures, 
$Y=\int n_{\mathrm{H}}n_{\mathrm{e}}\mathrm{d}V$, are given in 10$^{63}$~cm$^{-3}$. The scale factor $s$ is the ratio of the observed LSF to the expected LSF based on the overall radial surface brightness profile. The upper limits are quoted at their 95 per cent confidence level. Abundances are quoted with respect to the proto-solar values of \citet{lodders2003}. }
\begin{tabular}{lcccccc}
\hline
Parameter	&  5T-model  &   4-c.f. model \\
\hline
$Y_{\mathrm{0.25keV}}$	&  $<0.07$  	& --	\\
$Y_{\mathrm{0.5keV}}$	&  $0.79\pm0.05$& -- 	\\
$\dot{M}_{0.5-0.25keV}$ &	--	& $<0.06$ \\
$Y_{\mathrm{1.0keV}}$	&  $7.1\pm0.3$	& --	\\
$\dot{M}_{1.0-0.5keV}$ &	--	& $0.90\pm0.03$ \\
$Y_{\mathrm{2.0keV}}$	&  $56.3\pm1.3$ &   --    \\
$\dot{M}_{2.0-1.0keV}$ &	--	& $3.79\pm0.25$ \\
$Y_{\mathrm{3.0keV}}$	&  $<2.4$	& --	\\
$\dot{M}_{3.0-2.0keV}$ &	--	& $6.03\pm0.21$ \\
s		& $2.21\pm0.10$		& $2.19\pm0.11$	\\
C		& $0.75\pm0.21$		& $0.82\pm0.22$	\\
N		& $1.9\pm0.3$		& $2.0\pm0.3$	\\
O		& $0.79\pm0.04$		& $0.84\pm0.04$	\\
Ne		& $1.92\pm0.13$		& $1.92\pm0.13$	\\
Mg		& $1.25\pm0.15$		& $1.26\pm0.13$	\\
Fe		& $1.33\pm0.04$		& $1.39\pm0.05$	\\
Ni		& $0.9\pm0.3$		& $0.9\pm0.3$	\\
\hline
\label{RGStable}
\end{tabular}
\end{center}
\end{table}

\section{Discussion} 

\subsection{Orientation of the X-ray arms}

The inferred properties of the X-ray arms depend critically on their orientation with respect to our line-of-sight. 
Based on {\it HST} observations of the superluminal motion in the jet of M~87, \citet{biretta1999} concluded that the position angle of the jet is $<19^{\circ}$ from our line-of-sight. The two large, partially overlapping, 
outer radio lobes of M~87 \citep[the diffuse structures which embed the more collimated, brighter radio arms seen in the right panel of Fig.~\ref{M87im} and in][]{owen2000} are also probably oriented close to our line-of-sight. It is likely that the Southern radio lobe is positioned towards us and the Northern lobe is oriented away from us. This picture is supported by the observation of polarized flux from the Southern lobe \citep{andernach1979}, which shows that the Faraday depth towards it is likely to be small. The radius of the large lobes/bubbles is $\sim$22~kpc and their approximate centres are in projection 26~kpc apart. Assuming that the physical distance between the centres of the bubbles is at least twice their radius, a conservative limit for the orientation of the axis along which the bubbles rise is $<35^{\circ}$  from our line-of-sight. The Southwestern radio arm appears to ``merge'' into the Southern lobe indicating that its orientation is similar to that of the two large bubbles. Although the relation of the Eastern radio arm to the outer lobe is less clear, the most likely orientation of both X-ray and radio arms is approximately anti-parallel and far from the plane of the sky. 

\subsection{Physical properties and origin of the X-ray arms}

\begin{figure}
\includegraphics[width=1.1\columnwidth,clip=t,angle=0.,bb= 18 144 592 718]{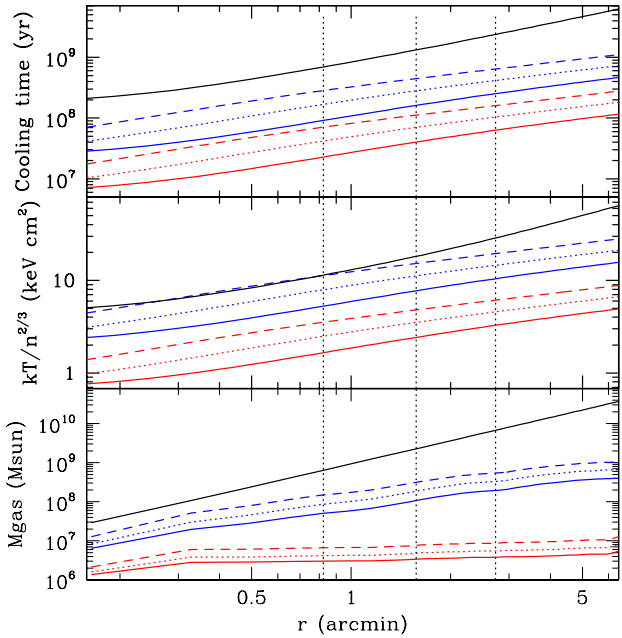}
\caption{Cumulative gas mass, entropy ($S=kT/n^{2/3}$), and radiative cooling time of the 0.5~keV phase (red lines), 1.0~keV phase (blue lines), and the ambient ICM (black lines) as a function of projected radius for 
orientation angles of 15$^{\circ}$ (dashed lines), 30$^{\circ}$ (dotted lines), and 90$^{\circ}$ (full lines) from our line-of-sight. For the deprojected density and temperature profiles of the ambient ICM we assume the profiles parametrized by equations (6) and (7) in \citet{churazov2008}. The cooler gas phases are assumed to be in pressure equilibrium with the ambient medium (see text for details). The vertical dotted lines denote the radii of the outer edge of the ridge of multiphase gas North of the jet, the end of the Southeastern horseshoe, and the radius of the shock at $\sim160$\arcsec\, which corresponds to the inner edge of the ``cap of the mushroom'' of the Eastern arm. } 
\label{profiles}
\end{figure}

We determine the properties of the X-ray arms for three different orientation angles (15$^{\circ}$, 30$^{\circ}$, and 90$^{\circ}$) from our line-of-sight. For each orientation, we determine the radial profiles of the gas mass, entropy, and cooling time for the different phases (see Fig.~\ref{profiles}). We assume that all the gas phases are in equilibrium with the 
ambient pressure at the given distance from the centre. The radial distribution of the ambient  pressure is determined using the deprojected temperature and electron density profiles parametrized by equations (6) 
and (7) in \citet{churazov2008}. Using this pressure profile $p(r)$, we calculate the electron number density profile $n_{\mathrm{e}}(r)=p(r)/(kT)$ of the $kT=0.5$~keV and 1~keV plasmas. Using these density 
profiles and the best fit emission measures, $Y=\int n_{\mathrm{H}}n_{\mathrm{e}}\mathrm{d}V$, for each extraction region, we determine the volumes which the cooler phases occupy: $V=Y/
(n_{\mathrm{H}}n_{\mathrm{e}})$, where $n_{\mathrm{H}}=n_{\mathrm{e}}/1.2$ is the hydrogen number density for collisionally ionized plasma with Solar metallicity. 

\begin{figure*}
\includegraphics[width=\textwidth,clip=t,angle=0.,bb=14 14 555 544]{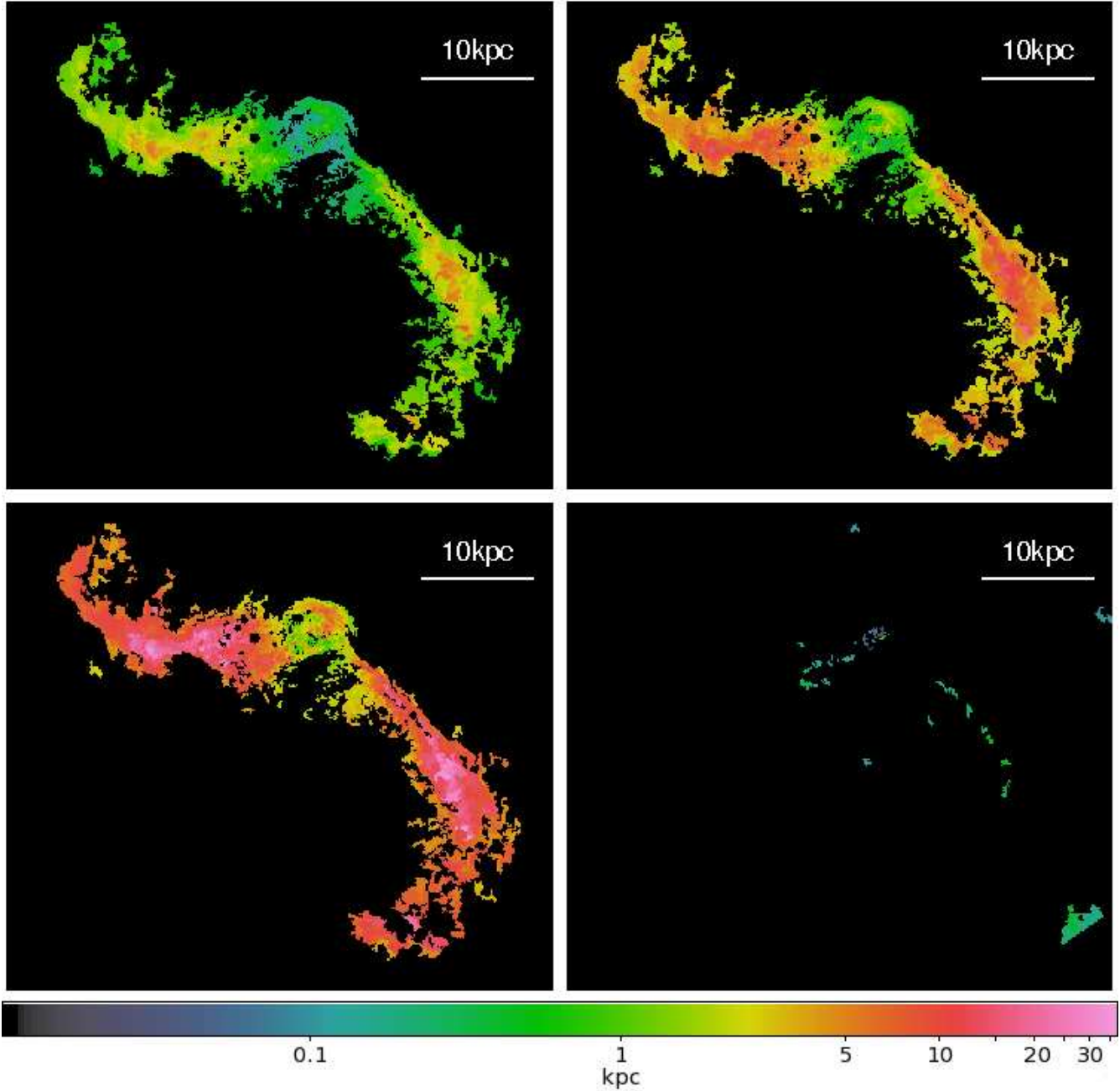}
\caption{Maps of the line-of-sight depths in kpc for the emitting volumes of 1~keV plasma for orientation angles of 90$^{\circ}$ (top left panel), 30$^{\circ}$ (top right panel), and 15$^{\circ}$ (bottom left panel) from 
our line-of-sight. The map on the bottom right panel shows the line-of-sight depth distribution of the 0.5~keV component for an orientation angle of 30$^{\circ}$ from our line-of-sight. All the panels show the same 
10.36\arcmin$\times$9.25\arcmin\ field. }
\label{fig:vff}
\end{figure*}

The line-of-sight depths, $l$, of the emitting volumes, $V$, shown in Fig.~\ref{fig:vff}, are then calculated as $l=V/A$, where $A$ is the area on the sky of the given extraction region. The corresponding gas masses are 
calculated as $M=1.4n_{\mathrm{H}}m_{\mathrm{p}}V$ (where $m_{\mathrm{p}}$ is the proton mass), only in the spatial regions where their presence was determined with better than 3$\sigma$ significance. The 
cumulative radial gas mass profiles are shown in the bottom panel of Fig.~\ref{profiles}. The middle panel of this figure shows the radial entropy profiles calculated as $S=kT/n_{\mathrm{e}}^{2/3}$. The top panel 
shows the cooling time profiles calculated as the gas enthalpy divided by the energy lost per unit volume of the plasma:
\begin{equation}
t_{\mathrm{cool}}=\frac{\frac{5}{2}(n_{\mathrm{e}}+n_{\mathrm{i}})kT}{n_{\mathrm{e}}n_{\mathrm{i}}\Lambda(T)},
\end{equation}
where the ion number density $n_{\mathrm{i}}=0.92n_{\mathrm{e}}$ and $\Lambda(T)$ is the cooling function for Solar metallicity tabulated by \citet{sutherland1993}. Cooling functions based on more up to date 
plasma codes \citep{schure2009} predict for the 0.5~keV plasma a 9 per cent higher cooling rate. 

If the arms lie in the plane of the sky, then the small line-of-sight depths of the emitting volumes in Fig.~\ref{fig:vff} indicate that the 1~keV gas is not volume filling at most radii, including the core of the galaxy. The 1~keV gas is then made of small blobs and filaments with an entropy smaller than the lowest entropy of the ambient medium in the centre of the galaxy (see Fig.~\ref{profiles}).
This would mean, as pointed out by  \citet{molendi2002}, that the cool blobs/filaments can hardly originate from the adiabatic evolution of the hot-phase gas entrained by buoyant radio bubbles, as suggested by \citet{churazov2001}. However, if the arms are oriented close to our line-of-sight, the physical picture changes. 

For the likely angles of 15$^{\circ}$--30$^{\circ}$ from our line-of-sight, both the radial distribution of the gas entropy and the distribution of the depths of the emitting volumes of the 1~keV component give a 
more consistent picture with other observables. The 1~keV plasma appears to be mostly volume filling, except in regions where the X-ray images reveal cavities. Its cooling time is  $>$$1.5\times10^8$~yr, which is comparable to or longer than the uplift time. Moreover, by taking the projection effects properly into account, the apparent large angle between the current direction of the jet and the Southwestern arm translates into a much smaller physical angle. The inferred spatial distribution of the depths of the emitting volumes (see Fig.~\ref{fig:vff}) indicates that the orientation angle of the filament is not constant and at certain positions the arms are folded along our line-of-sight. The Southwestern arm appears to be folded at the projected radius of 3.4\arcmin, where the arm divides into two filaments (see Fig.~\ref{M87im}). The Eastern arm appears to have folds at $\sim$2.1\arcmin\ and its line-of-sight depth increases in the cap of the mushroom. We caution that because of the strongly simplified assumptions the values for the depths quoted in the figures are approximate numbers only.

If the uplift of the low entropy plasma were strictly adiabatic, then the temperature of the arms would decrease as a function of radius. However, as shown by \citet{molendi2002}, \citet{simionescu2008a}, and in Fig.~\ref{kTprof} using a two-temperature model, the average temperature of the arms is remarkably constant. The constant temperature implies rising entropy with radius and suggests that conduction plays an important role in the thermal evolution of the arms. 

The estimated total mass of 1~keV gas in the arms is between $6$--$9\times10^8$~$M_{\odot}$, which given the uncertainties is remarkably similar to the total gas mass of $6.3\times10^8~M_{\odot}$ within the innermost 3.8~kpc region (0.8\arcmin\ in the plane of the sky), where the X-ray emission is relatively spherically symmetric. Outside of this radius, all the $\sim$1~keV gas is distributed along the two X-ray arms.  
If the entropy of the arms is increasing due to heat conduction, then most of the 6--9$\times10^8~M_{\odot}$ of $\sim$1~keV plasma might have been uplifted from within the innermost approximately spherical 3.8~kpc region of the galaxy. While this is a large fraction of the total gas mass in the core of the cluster, it is only a fraction of the uplifted gas mass observed in more massive clusters; e.g. the inferred mass of the uplifted gas in Hydra A is 1.6--$6.1\times10^9~M_{\odot}$ \citep{simionescu2009b}. Though seen with great clarity in our study, the physics of gas uplift in M~87 is by no means extreme. Many brightest cluster galaxies can be expected to be similarly efficient in evacuating their lowest entropy plasma during AGN outbursts. 

By assuming that the outer radio halo is a spherical bubble of hot plasma undergoing a steady energy input from the jet, \citet{owen2000} estimated its age to $\sim$$10^8$~yr. 
However, if the radio halo consists of two bubbles rising buoyantly in the hot ICM, then the outer radio lobes are likely to be older. The age of the radio and X-ray arms estimated by 
\citet{churazov2001} is a few times 10$^7$~yr. However, if the whole Southwestern arm originated in a single uplift following an AGN outburst, then for an orientation angle of $<$$30^{\circ}$ from our line-of-sight, its length is $>$60~kpc, which assuming an uplift velocity of 400~km~s$^{-1}$ implies an age of $>$$1.5\times10^8$~yr. The age of the observed structures is thus greater than previously thought.

\subsection{Implications for gas motions in the ICM}

The NW edge of the Southwestern X-ray arm is remarkably smooth out to the radius of $\sim$3.3\arcmin\ and the filament is surprisingly straight for a distance of over 2\arcmin. This strongly suggests that the gas motions cannot be very turbulent in either phase. The lack of strong turbulence in the ICM has been inferred previously based on the straightness of the H$\alpha$ filaments in the Perseus cluster \citep{fabian2003b}, and strong upper limits on turbulent velocities were placed based on resonance scattering in NGC~4636 \citep{werner2009} and turbulent line broadening in Abell~1835 \citep{sanders2009c}. The bending of the X-ray arms of M~87 at larger radii might be caused by gas sloshing, which creates a spiral like shearing motion pattern in the ICM which is the most obvious at the two opposite and staggered cold fronts seen at 7\arcmin\ to the Southeast \citep[][]{simionescu2007} and at 19\arcmin\ to the North \citep{simionescu2010}. The sloshing induced shearing motions might also generate mixing of the $\sim$1~keV gas with the ambient ICM at large radii.

\subsection{The coolest X-ray emitting phase}

The total mass of $\sim$0.5~keV gas in M~87 is at least $7\times10^{6}~M_{\odot}$. 
Because we calculate the mass of the 0.5~keV component only in spatial regions where it was detected at higher than the 99.7 per cent confidence, this value is a conservative lower limit. The inferred small line-of-sight depths of the emitting volumes of the 0.5~keV component imply that the coolest X-ray emitting gas is not volume filling, but forms a multi-phase medium together with both the hotter X-ray emitting and the cold 
H$\alpha$ emitting gas phases. Soft band X-ray images \citep{forman2007} reveal that this cooler phase forms a network of filaments, which spatially coincide with $H\alpha$ filaments \citep[][see also Fig.~\ref{Halpha}]{sparks2004}. Relatively cool ($\sim0.5$~keV) X-ray gas associated with H$\alpha$ filaments has also been observed in the Perseus cluster and in 2A~0335+096 \citep{sanders2007,sanders2009}. 

One of the most prominent regions in M~87 occupied by multiphase plasma is a ridge to the North of the jet, which contains 2~keV, 1~keV, 0.5~keV, and H$\alpha$ gas. This region is just outside of the inner radio cocoon and possibly contributes to the confinement of the radio plasma on the Northern side and to the depolarization seen there by \citet{owen1990}.  The depolarization of the radio emission in this region overlapping with the filaments indicates that the magnetic fields are disordered on scales less than 0.1~kpc. Therefore, it is likely that the H$\alpha$ filaments consist of many narrow magnetized threads, which are not perfectly aligned and are small compared to the observed beam. Depolarization of the radio emission at regions containing H$\alpha$ and soft X-ray emission has been previously observed in the Centaurus cluster \citep{taylor2007}.  

\subsubsection{Limits on cooling and constraints on viable heating mechanisms}

Even though the cooling time of the $\sim$0.5~keV gas is relatively short (see the top panel of Fig.~\ref{profiles}), no X-ray emitting plasma below 0.5~keV is observed in M~87. The 95 per cent confidence upper limit for the 
emission measure of 0.25~keV plasma based on the non-detection of \ion{O}{vii} lines with RGS is $Y=\int n_{\mathrm{H}}n_{\mathrm{e}}\mathrm{d}V=7\times10^{61}$~cm$^{-3}$, which is only 9 per cent of the 
emission measure of the 0.5~keV plasma. Therefore, assuming pressure equilibrium between different gas phases, the mass of the 0.25~keV phase is at least 22 times lower than the mass of the 0.5~keV plasma. 
Assuming the 1~keV gas is isobarically cooling to 0.5~keV, its best fit mass deposition rate is $\dot{M}=0.9~M_{\odot}$~yr$^{-1}$, which is 15 times larger than the 95 per cent confidence upper limit for the mass deposition rate 
from 0.5~keV to 0.25~keV. The upper limits on gas cooling below 0.5~keV rely on the \ion{O}{vii} lines and thus on the O abundance of the cool gas, which we assumed to be the same as that of the hotter phase which is fairly constant across the core \citep{werner2006b}. For significant amounts of gas cooling radiatively below 0.5~keV to be present in the core, without detectable \ion{O}{vii} emission, the O abundance of the cooling plasma would have to be unrealistically low. 
The lack of larger amounts of gas radiatively cooling below 0.5~keV implies the presence of heating. A similar temperature floor at $\sim$0.5~keV is observed in the Perseus cluster, 
Centaurus cluster, 2A~0335+096, Abell 262, Abell 3581, and HCG 62, \citep{sanders2007,sanders2008,sanders2009,sanders2009b}.  

The energy emitted per unit volume in a blob of 0.5~keV gas, that is in pressure equilibrium with the surrounding 1~keV and 2~keV gas, is 4.8 times larger than the energy emitted by the neighboring volume of gas 
at 1~keV and 30 times larger than for the gas at 2~keV. Therefore, any heating process preventing the coolest gas phase from cooling below the $\sim$0.5~keV temperature floor and providing energy uniformally 
in a volume-averaged way would overheat the surrounding gas and is ruled out by the data. The mechanism that heats the coolest gas clearly provides more energy per unit volume to cooler, denser phases than it 
does to the hotter phases. 

An alternative explanation is that the coolest X-ray plasma cools non-radiatively by mixing. \citet{sanders2009b} reached similar conclusions for a sample of cool core clusters. 
The coolest X-ray plasma in M~87 might cool non-radiatively by mixing with the cold optical line emitting filamentary gas, allowing efficient conduction between the two phases  \citep{fabian2002,soker2004}. As 
indicated by the fact that all the bright H$\alpha$+[\ion{N}{ii}] filaments end at the innermost shock front, this mixing might be promoted by the shock, which induces enhanced shearing motions, and might also power the detected optical and UV line emission (see the next section).

\subsection{The nature of the H$\alpha$ filaments}

The optical emission line filaments in M~87 are likely due to cooled gas, enriched by dust from stellar mass loss \citep{sparks1993}, which has been uplifted by the buoyant relativistic plasma, together with the 
coolest X-ray emitting phase. The total mass of the $\sim$$10^4$~K H$\alpha$ emitting gas in the filament system in M~87 is estimated to be of the order of $10^5$--$10^7~M_{\odot}$ \citep{sparks1993}, which is comparable to the mass of $\sim$0.5~keV gas. However, the total mass of cold neutral and molecular gas based on the upper limits on CO emission within the bright H$\alpha$ emitting regions is only a few times $10^6~M_{\odot}$ \citep{salome2008}. This implies that the H$\alpha$ emitting gas is not a thin skin on the underlying clouds of neutral and molecular gas as in the Perseus cluster, where the inferred total mass of the cold gas is as high as $\sim$4$\times10^{10}~M_{\odot}$ \citep{salome2006}.  Recently, \citet{sparks2009} discovered filaments of \ion{C}{iv} emission in M~87, with a total power of $1.1\times10^{39}$~erg~s$^{-1}$, arising from gas at a temperature of $\sim$$10^5$~K, which spatially coincides with the H$\alpha$ filaments on all scales. This detection indicates that the cool and hot gas phases are in thermal communication. The emission strengths are consistent with thermal 
conduction, which would create this intermediate temperature phase at the interface of the hot ICM and the cooler gas. However, even though conduction can quantitatively provide the energy seen in the form of 
line emission \citep{sparks2004}, it would have to proceed at close to the Spitzer or saturated levels to power the filaments. The very thin and long H$\alpha$ filaments are magnetized \citep[see Sect. 4.2 and][]{fabian2008} and the likely orientation of the magnetic fields parallel to the interface of the cold and hot phase would largely suppress the level of conduction. Thus it seems unlikely that thermal conduction alone can explain the optical 
line emission.

All of the bright H$\alpha$ emission in M~87 is observed in the downstream region of the innermost $<$3~Myr old shock front, within a radius of $\sim0.6$\arcmin. This suggests that the H$\alpha$ emission in M~87 is somehow related to shocks in the ICM. The passage of a shock accelerates the intra-cluster medium, but because of the large density contrast the relatively cool H$\alpha$ emitting gas is left behind, resulting in a sudden strong shear around the filaments. It is likely that such shear promotes mixing of the cold gas with the ambient hot ICM via instabilities. 
 \citet{ferland2009} recently showed that the spectra of optical filaments observed around the central galaxies of cooling core clusters can be explained if the filaments are heated by ionizing particles, either 
conducted from the surrounding regions or produced in situ by MHD waves. By facilitating contact between the hot thermal particles and the cold gas, mixing can supply the power and the ionizing particles needed to explain the optical line emission spectrum.  Mixing of X-ray emitting plasma with the cold gas also naturally explains the presence of the $\sim10^5$~K intermediate temperature gas phase. 
If the coolest X-ray emitting gas cools by mixing with the filamentary gas behind the shock front and conductively connects with it \citep[e.g.][]{begelman1990,soker2004}, then its thermal energy will be 
radiated away in the UV band, in the optical H$\alpha$+[\ion{N}{ii}], and most likely at infrared wavelengths in the presence of dust.

\subsection{Implications for galaxy formation models}

The series of recent `radio' or `jet'-mode AGN outbursts in M~87, revealed in exquisite detail by the \chandra\ data, by no means represent a unique or extreme phenomenon. For example, a complete, flux-limited study of eighteen optically and X-ray bright, nearby galaxies, including M~87, by \citet{dunn2010} shows that 17/18 sources exhibit some form of current radio activity associated with the central AGN \citep[see also][]{best2005,dunn2006,dunn2008,sun2009}. \citet{allen2006} also show that the efficiency with which the accretion of hot gas powers jets in giant ellipticals, including M~87, is remarkably uniform. Our results for the nearest, brightest cluster core show that even the current, relatively modest level of radio mode AGN activity has been sufficient to remove a sizeable fraction of the coolest, lowest entropy gas from the central galaxy. This is the material that would have cooled to form stars \citep{peterson2006}. Our results therefore provide direct evidence that radio mode AGN input has a profound effect on the ability of large galaxies to form stars from their surrounding X-ray emitting halos. 

\section{Conclusions}

Using a combination of deep (574~ks) \chandra\ data, \xmm\ high-resolution spectra, and optical H$\alpha$+[\ion{N}{ii}] images, we have performed an in-depth study of AGN feedback in M~87. Accounting for the 
fact that the long X-ray and radio ``arms'' are seen in projection with likely angles of $\sim15^{\circ}$--$30^{\circ}$ from our line-of-sight, we find that:

\begin{itemize}

\item 
The mass of the uplifted low entropy gas in the arms is comparable to the gas mass in the approximately spherically symmetric 3.8~kpc core, demonstrating that the AGN has a profound effect on its immediate surroundings. This result has important implications for understanding AGN feedback in galaxy formation models \citep[e.g.][]{croton2006,delucia2007,sijacki2007}. 

\item 
The coolest detected X-ray plasma, with a temperature of $\sim$0.5~keV, the H$\alpha$+[\ion{N}{ii}] emitting gas, and the UV emitting gas are cospatial, and appear to be arranged in magnetized filaments within 
the hot ambient atmosphere, forming a multi-phase medium.

\item  
We do not detect X-ray emitting gas with temperatures below $\sim$0.5~keV. To remain in steady state and at constant pressure, the filaments of 0.5~keV gas would require a 5 times higher heating rate than the 
1~keV gas, and 30 times stronger heating than 2~keV plasma. Thus, a uniform, volume-averaged heating mechanism would overheat the surrounding gas and can be ruled out in M~87.  

\item 
All of the bright H$\alpha$ and UV filaments are seen in the downstream region of the $<3$~Myr old shock front, within a radius of  $\sim$0.6\arcmin. This argues that the generation of H$\alpha$ and UV 
emission in M~87 is related to shocks in the ICM. 
\end{itemize}

Based on these observations we propose that shocks induce shearing around the denser gas filaments, thereby promoting mixing of the cold gas with the ambient hot ICM via instabilities.  This process may both 
provide a mechanism for powering the optical line emission and explain the lack of plasma with temperature under 0.5~keV: 

\begin{itemize}
\item
 By helping to get hot thermal particles into contact with the cool, line-emitting gas, mixing can supply the power and the ionizing particles needed to explain the optical spectra of the H$\alpha$+[\ion{N}{ii}] nebulae 
\citep[see][]{ferland2009}. 

\item
Mixing of the coolest X-ray emitting plasma with the cold optical line emitting filamentary gas promotes efficient conduction between the two phases, allowing the X-ray gas to cool non-radiatively, which may explain 
the lack of gas with temperatures under 0.5~keV. 
\end{itemize}

\section*{Acknowledgments}
Support for this work was provided by the National Aeronautics and Space Administration through Chandra/Einstein Postdoctoral Fellowship Award Number PF8-90056 and PF9-00070 issued by the Chandra X-
ray Observatory Center, which is operated by the Smithsonian Astrophysical Observatory for and on behalf of the National Aeronautics and Space Administration under contract NAS8-03060. This work was 
supported in part by the US Department of Energy under contract number DE-AC02-76SF00515. SMH is supported by the National Science Foundation Postdoctoral Fellowship program under award number AST-0902010.

\bibliographystyle{aa}
\bibliography{clusters}

\end{document}